\documentclass[aps,pra,twocolumn,superscriptaddress,floatfix,longbibliography]{revtex4-1}

\usepackage{amssymb}
\usepackage{graphicx}
\usepackage{dcolumn}
\usepackage{bm}
\usepackage{amsmath}
\usepackage{ulem}
\usepackage[colorlinks,linkcolor=magenta,citecolor=blue,urlcolor=blue]{hyperref}

\begin{document}

\title{Three-body bound states of two bosons and one impurity in one dimension}
\author{Yanxia Liu}
\affiliation{Beijing National Laboratory for Condensed Matter Physics, Institute of Physics, 
Chinese Academy of Sciences, Beijing 100190, China}
\author{Yi-Cong Yu}
\email{ycyu@wipm.ac.cn}
\affiliation{Beijing National Laboratory for Condensed Matter Physics, Institute of Physics, 
Chinese Academy of Sciences, Beijing 100190, China}
\affiliation{State Key Laboratory of Magnetic Resonance and Atomic and molecular Physics,
Wuhan Institute of Physics and Mathematics, Innovation Academy for Precision Measurement
Science and Technology, Chinese Academy of Sciences, Wuhan 430071, China}

\author{Shu Chen}
\email{schen@iphy.ac.cn}
\affiliation{Beijing National Laboratory for Condensed Matter Physics, Institute of Physics, 
Chinese Academy of Sciences, Beijing 100190, China}
\affiliation{School of Physical Sciences, University of Chinese Academy of Sciences, Beijing, 100049, China}
\affiliation{Yangtze River Delta Physics Research Center, Liyang, Jiangsu 213300, China}

\begin{abstract}
We investigate one-dimensional three-body systems composed of two identical bosons and one mass-imbalanced atom (impurity)
with attractive two-body and three-body zero-range interactions. In the absence of three-body interaction,
we give a complete phase diagram of the number of three-body bound states in the whole region of mass ratio and the ratio 
of intra- and inter-component interaction strength via direct calculation of Skornyakov-Ter-Martirosyan
equations. We demonstrate that other
low-lying three-body bound states emerge when the mass of the impurity particle is different from other two identical particles. 
We obtain the binding energies together with the corresponding wave functions. When the mass of impurity atom is 
vary large, there are at most three three-body bound states. 
In the presence of three-body zero-range interaction, we unveil that 
weak three-body interaction will not always induce one more three-body bound state. At some special parameter points, 
arbitrary small  three-body interaction can generate one more three-body bound state. This corresponds to the transition 
of the number of three-body bound states induced only by two-body attractive interaction.

 \end{abstract}

\maketitle

\section{introduction}

The  quantum three-body problem has drawn numerous concerns and is of central interest in the study of few-body
physics \cite{PhysRep,PhysRep2,RMP,RPP}. In the past decades, continuous efforts have led to many theoretical
breakthroughs \cite{Skorniakov1957,Fad63,Efimov1970,Efimov1970b,Efimov1971}, including
the derivation of the well-known Skornyakov-Ter-Martirosyan (STM) equations which can be used to calculate
the wave functions and spectra of quantum identical three-body system with short-range interactions \cite{Skorniakov1957},
the Faddeev's formulism  of three-body problem with discrete and continuum spectrum \cite{Fad63}, and
the finding of distinctive Efimov effect in the spectrum and trimer states of the three-boson
system \cite{Efimov1970,Efimov1970b,Efimov1971}.
The three-boson system with short-range interactions can exhibit infinite number of trimer states
fulfilling a discrete symmetry, which is named as the Efimov effect. The first experimental
evidence of the Efimov effect came from ultracold gases \cite{Kraemer2006}, and this early
evidence stimulated intensive studies of few-body ultracold physics in different
dimensions
\cite{Braaten2007,Naidon2017,Ferlaino2010,Zaccanti2009,Nishida2013,Kunitski2015,Mattis1986,
Shi2014,Cui2014,Guan2018,Nishida2013b,Moroz2015}.
Experiments with few cold atoms provide unprecedented control on both the
atom number with unit precision and the interatomic interaction strength
by combination of sweeping a magnetic offset field and the confinement
induced resonance \cite{Chin2010}.

Recently, three-body systems in one dimension (1D) have gained a lot of attention
\cite{Mora2005,Kartavtsev2009,Mehta2014,Mehta2015,Moroz2015,
Nishida2018,Guijarro2018,Pricoupenko2019,Harshman2020,Happ2019}. As the basis of quantum integrability,
the Yang-Baxter equation describes the two-body scattering
matrix fulfilling a certain intertwined relation with at least three particles, and thus three-body systems become
important candidates in studying the integrability and its breakdown \cite{Mazets2008,Petrov2012,
Kristensen2016,Tan2010,Lamacraft2013}. An integrable three-boson system with two-body attractive interactions 
is known to have only a three-body bound state \cite{LL,McGuire}. In general, the introduction of mass imbalance 
and three-body interaction will break the integrability condition. Nevertheless, it has been shown that the 
imbalanced three-body systems exhibit more rich physics than the integrable systems which are composed 
of three identical atoms \cite{Kartavtsev2009,Mehta2014,Mehta2015,Happ2019}. 
The zero-range three-body forces in quasi-1D system can be induced by the virtual excitations of
pairs of atoms in the waveguide \cite{Mazets2008,Tan2010}, which may realize the quantum droplets
in one dimensional system \cite{Sekino2018,Pricoupenko2018a}.
For 1D interaction systems, some physical properties will not disappear 
in the presence of three-body interaction, for example, Bose-Fermi mapping \cite{Girardeau1960, 
Cheon1999,Girardeau2004,Valiente2020,Sekino2021,Valiente2021}.

For the system of three identical particles with attractive three-body interaction, there 
exists an excited trimer state in the vicinity of
the dimer threshold \cite{Guijarro2018,Nishida2018}.

Most of the theoretical studies on  mass-imbalanced systems in 1D  focused on
the heavy-heavy-light (HHL) system \cite{Kartavtsev2009,Mehta2014,Mehta2015,Happ2019} in
which case the Born-Oppenheimer approximation (BOA) and the adiabatic
hyperspherical approximation work relatively well. This system has a rich three-body 
bound state spectrum and the number of bound states increases with increasing 
heavy-light mass ratio. Meanwhile, the experimental realizations
of mass-imbalanced systems have made tremendous progresses,
such as fermionic mixtures \cite{Wille2008,Tiecke2010,Cetina2016,Ravensbergen2018}
and bosonic-fermionic mixtures \cite{Hadzibabic2002,Guenter2006,Best2009,Wu2011,Tung2014,Lous2018},
which stimulate us to theoretically investigate mass-imbalanced three-body systems in the whole 
parameter regions and beyond the BOA.

In this work, we study 1D three-body systems composed of two identical bosons and one impurity
with zero-range two-body and three-body interactions by solving the momentum-space STM
equations. We first study the case in the absence of three-body interaction and present 
the phase diagram of the number of bound states in the parameter space spanned by the mass 
ratio and ratio of intra- and inter-component interaction strength. We find 
significant differences between light-light-heavy (LLH) and HHL systems. Particularly, 
we unveiled that the LLH system possess at most three three-body bound states 
with attractive interactions.
We then study the effect of three-body zero-range interaction and derive the corresponding STM equations. 
At some special parameter points, one more three-body bound state induced 
by three-body interaction for arbitrary strength comes into presence, 
compared with the cases only with two-body attractive interaction. 
These points correspond to the transition points of the number of three-body bound states induced only 
by two-body attractive interaction.

Our article is organized as follows. In Sec. II we first introduce our model and then describe the method
for solving our three-body problem in details.
Particularly, we develop some computational techniques to calculate the STM equation by mapping it to
solving linear equations which enables us to get the complete phase diagram of the number
of three-body bound states in the whole parameter region, which is shown in Sec. III. We also present the
exact Bethe-ansatz solution of the odd-parity bound state in the limit case where the impurity is infinitely heavy.
In Sec. IV, the three-body interaction is introduced and we show how the mass ratio as well as the ratio of coupling strengths
effect the forming of three-body bound state induced by the three-body interaction. A summary is given in Sec. V.

\section{The model and method} \label{secMM}
\subsection{The model}
The general Hamiltonian for a three-particle system composed of two identical bosons (1 and 2) with
mass $M$ and an impurity particle (3) with mass $m$ in
one dimension \cite{Kartavtsev2009} is given by
\begin{align}
\hat{H} = &- \frac{\hbar^2}{2M}( \frac{\partial^2}{\partial x_1^2}+\frac{\partial^2}{\partial x_2^2})
-\frac{\hbar^2}{2m}\frac{\partial^2}{\partial x_3^2}+d_0\delta(x_1-x_2)\notag \\
&+g_0\delta(x_1-x_3)+g_0\delta(x_2-x_3),
 \label{eqHamiltonian}
\end{align}
where the attractive boson-boson (BB) and boson-impurity (BI) interactions
are described by zero-range $\delta$-functions with coupling constants $d_0<0$ and $g_0<0$.
Note that the total momentum $\hat{P} = \sum_i -\mathrm{i}\hbar \frac{\partial}{\partial x_i} $
is conserved. Thus by introducing the Jacobi coordinates:
\begin{align}
x&=x_3-\frac{x_1+x_2}{2}, \quad
 y&=\frac{\sqrt{2M/m+1}}{2}(x_1-x_2),
\label{eqCoordinateTransformation}
\end{align}
the time-independent Schr\"odinger equation of (\ref{eqHamiltonian}) with
its binding energy $E= -\hbar^2\kappa^2/(2\mu_{12,3})$
can be reduced in the center-of-mass frame as
\begin{equation}
 \begin{split}
& [-(\frac{\partial^2}{\partial x^2} +\frac{\partial^2}{\partial y^2} )
+g\delta(x\sin\theta-y\cos\theta) \\
&+ g\delta(x\sin\theta+y\cos\theta) + d\delta(y)+\kappa^2 ]\psi=0,
\label{eqHamiltonian2D}
\end{split}
\end{equation}
\begin{figure}[htbp]
\centering
\includegraphics[width=8.5cm]{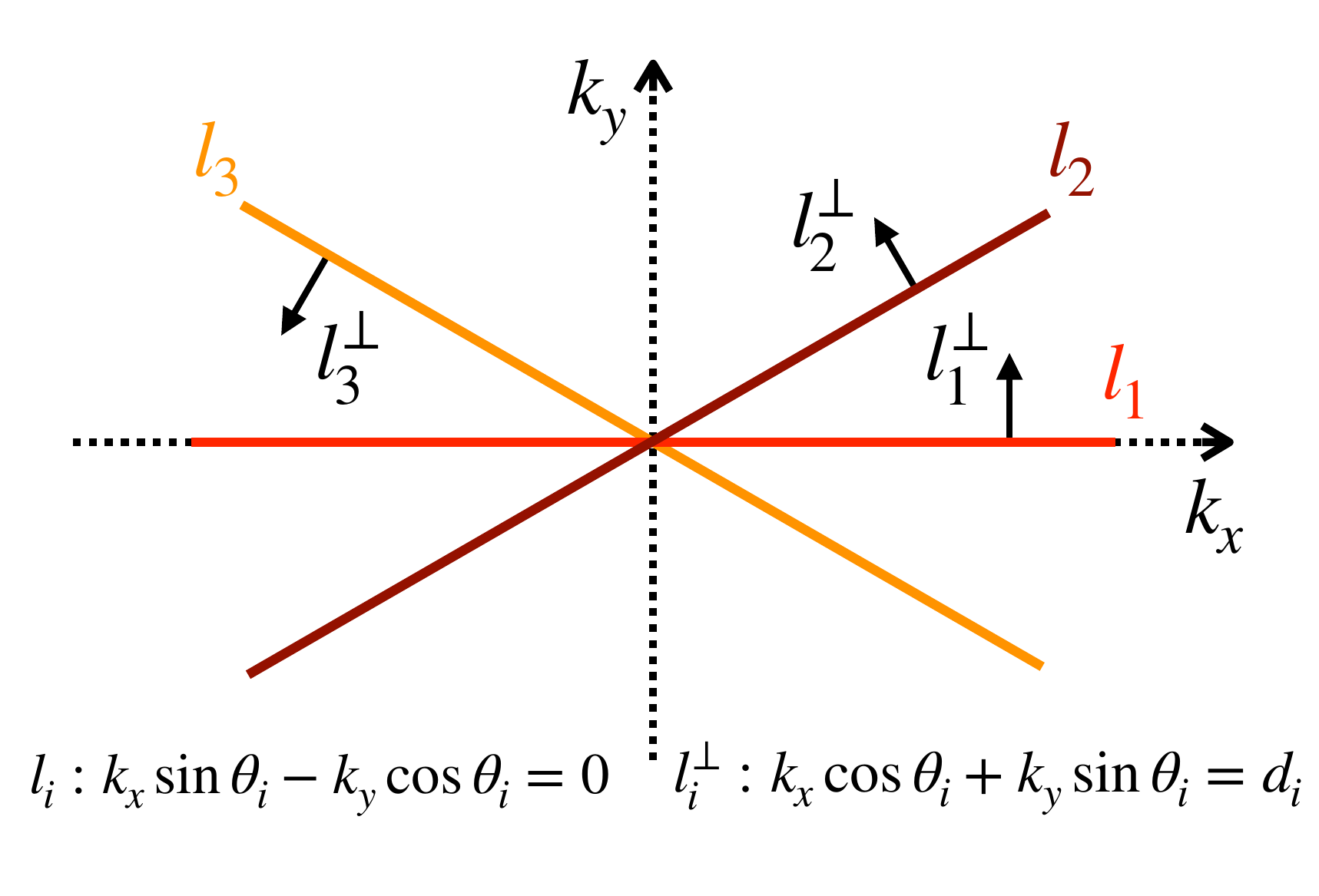}
\caption{Schematic of the Hamiltonian (\ref{eqHamiltonian2D}).The solid lines
presents the delta potentials.}
\label{fig1}
\end{figure}
where $\mu_{12,3}=2Mm/(2M+m)$, $\theta = \arctan\sqrt{1+2M/m}$ and the rescaling coupling constants
are
\begin{align}
d &= d_0\frac{\mu_{12,3}}{\hbar^2}\sqrt{1+2M/m}=-\frac{4}{\tan\theta a_{BB }}, \notag\\
g &= g_0\frac{2\mu_{12,3}}{\hbar^2}\sqrt{\frac{1+2M/m}{2+2M/m}}=-\frac{2\sqrt{2+2M/m}}{\tan\theta a_{BI}}.
 \label{eqGdtheta}
\end{align}
The BB and BI scattering lengthes are $a_{BB}=-\hbar^2/(\mu_{BB}d_0)$ and
$a_{BI}=-\hbar^2/(\mu_{BI}g_0)$ with $\mu_{BB}=M/2$ and $\mu_{BI}=Mm/(M+m)$. The limitation $M/m \to \infty$ and $M/m \to 0$ are represented
by $\theta \to \pi/2$ and $\theta \to \pi/4$ respectively. And $\theta = \pi/3$ corresponds to the
equal-mass case.
%
\subsection{The STM equations}
The Hamiltonian (\ref{eqHamiltonian}) was investigated by several articles
\cite{Kartavtsev2009,Happ2019} with $M/m>1$ and is of great interests in recent experiments \cite{Zundel2019}. It has been confirmed that in the limit $|g_0/d_0| \to 0 $
and $|g_0/d_0| \to \infty$, there exists a critical value of the mass ratio $M/m$  where three-body bound states emerge and
the (2+1)-scattering length vanishes. However, previous studies relied  on the BOA
in strong (weak) coupling limit on the premise of  $M/m \gg 1$ \cite{Born1927}.
The BOA is not proper near $M/m=1$ or $M/m<1$, and will lead to the loss of important information on the bound states.

In this work, we adopt the method developed in \cite{Nishida2018,Guijarro2018}
to transform the  Hamiltonian (\ref{eqHamiltonian2D}) into three coupled integral equations. Solving
these integral equations  provides
the wave functions of bound states and eigen energies
and further gives the full phase diagram of the number of bound states.
The time-independent Schr\"odinger equation (\ref{eqHamiltonian2D})
in momentum space reads
\begin{align}
(p_x^2+p_y^2+\kappa^2) u(p_x,p_y) + \sum_{i=1}^3 \frac{g_i}{2\pi} \int \mathrm{d} l_i^\bot u(k_x,k_y) = 0,
\label{eqSchEq}
\end{align}
where $g_1=d$, $g_2=g_3=g$, $E$ is the eigenenergy,
and $u(p_x,p_y)$ is the wave function in momentum space
 $u(p_x,p_y) = \int \frac{\mathrm{d}x\mathrm{d}y}{2\pi}
\psi(x,y) \mathrm{e}^{\mathrm{i}(-p_x x - p_y y)}$.
The $\mathrm{d} l_i^\bot$ denotes the line integral of the complex scalar
field $u(k_x,k_y)$ with the path parameterized by
$k_x = d_i\cos\theta_i - t_i \sin\theta_i$, $k_y = d_i \sin\theta_i + t_i\cos\theta_i$, 
where $t_i=-k_x\sin\theta_i + k_y \cos\theta_i$ is the arc length parameter  and
$d_i = p_x\cos\theta_i + p_y \sin\theta_i$, representing the integral line
$l_i^\bot$ that goes through the point $(p_x,p_y)$ and being vertical to line $l_i$,
see Fig.\ref{fig1}.
Here $\theta_1 = 0$, $\theta_2 = \theta$, $\theta_3 = -\theta$, and $\theta$ solely depends
on the mass ratio
$M/m$, see Eq. (\ref{eqGdtheta}).
Note that after integrating along line $l_i^\bot$, the results can be arranged as a one-parameter function
$
f_i(d_i) = \int \mathrm{d} l_i^\bot u(k_x,k_y),
$
and the Schr\"odinger equation (\ref{eqSchEq}) becomes
\begin{align}
(p_x^2+p_y^2+\kappa^2) u(p_x,p_y) + \sum_{i=1}^3 \frac{g_i}{2\pi} f_i(d_i) = 0.
\label{eqSchEq2}
\end{align}
The integration of Eq. (\ref{eqSchEq2}) over $l_i^{\bot}$ leads to
\begin{align}
&\left(1+\frac{g_i}{2\sqrt{k^2+\kappa^2}} \right)f_i(k) \notag \\
&= \sum_{j \neq i} \int \frac{\mathrm{d} k'}{2\pi} \frac{- g_j \vert \sin(\theta_i - \theta_j)\vert f_j(k')}
{k'^2+k^2-2 k k'\cos(\theta_i-\theta_j) +\kappa^2\sin^2(\theta_i-\theta_j)}
\label{eqSTM}
\end{align}
for $i = 1,2,3$, which are the STM equations in momentum space.
Substituting $f_i(k)$ into (\ref{eqSchEq2}) and then  taking
Fourier transformation, the solutions of the Shr\"odinger equation (\ref{eqHamiltonian2D})
are obtained.

Two remarks are necessary for solving (\ref{eqSTM}): first, the self-consistent conditions
$\int \mathrm{d}k_x \mathrm{d}k_y u(k_x, k_y) = \int \mathrm{d}k f_i(k) $ can be proved by 
integrating (\ref{eqSTM}) by both sides;
second, the Hamiltonian (\ref{eqHamiltonian2D}) obviously  possesses exchange symmetry of
two bosons and parity symmetry, which are reflected in the eigenstates of  (\ref{eqHamiltonian2D})
by $\psi (x,y)= \psi(x,-y)$
and $\psi(x, y)= \pm \psi(-x,-y)$, and these two
discrete symmetries are well represented in (\ref{eqSTM})
by $f_2(k) =  f_3(k)$ for $\psi (x,y) = \psi(x,-y)$
and $f_i(k) = \pm f_i(-k),i=1,2,3$ for $\psi(x,y) = \pm \psi(-x,-y)$, respectively.
%

The solutions $f_i(k)$ of STM equations (\ref{eqSTM}) have no pole in bound state sector, thus 
$f_i(k)$ can be safely discretized numerically.
The analysis of the
scattering sector by (\ref{eqSTM}) is much more sophisticated because the singularities
of wave function $u(p_x,p_y)$ need careful handling. In this work we concentrate on 
the bound states sector only.

After discretization, the STM equations (\ref{eqSTM})
becomes a linear equation set and the non-zero solutions
satisfying $E<E_{\text{th}}$ are the bound states. Here $E_{\text{th}} = -\hbar^2\max\{(g_i/2)^2\}/(2\mu_{12,3})$
denotes the two-body threshold energy and serves as
the lower bound of the continuous spectra.
Specifically, consider the combination of
the three functions $f_1(k)$, $f_2(k)$, $f_3(k)$ as a vector $[f_1(k),~f_2(k),~f_3(k)] \to \bm{V}$,
then the three STM equations (\ref{eqSTM}) become a matrix equation
$
\bm{M}(E)\bm{V} = \bm{V}.
$
The existence of non-zero solution of equation
$\text{det}[\bm{M}(E)-I]=0$ gives spectrum of the Hamiltonian.

For given $g_i$ and $\theta_i$, to obtain non-zero solution $f_i(k)$ in (\ref{eqSTM}),  we need to search for the 
discrete energies $-\kappa^2$, which is a difficult task.
However,  we can bypass this difficulty by solving the eigenvalue problem
\begin{align}
&\left(-\lambda+\frac{g_i}{2\sqrt{x^2+\tilde{\kappa}^2}} \right)\tilde{f}_i(x) \notag \\
&= \sum_{j \neq i} \int \frac{\mathrm{d} y}{2\pi} \frac{-g_j \vert \sin(\theta_i - \theta_j)\vert \tilde{f}_j(y)}
{y^2+x^2-2 x y\cos(\theta_i-\theta_j) +\tilde{\kappa}^2\sin^2(\theta_i-\theta_j)},
\label{eqSTMeig}
\end{align}
where $\lambda$, which can be numerically proved negative definite, 
in the LHS of Eq. (\ref{eqSTMeig}) is the eigenvalue and $\tilde{\kappa}$ is set to be a unit whose 
value can be arbitrarily chosen (for convenience, we set $\tilde{\kappa}=1$). 
In this sense, solving $\lambda$ gives us the solutions of 
Eq. \eqref{eqSTM} as $\kappa=\lambda\tilde{\kappa}$. The functions $f_j(k)$ can be obtained by 
taking $f_j(k)= \tilde{f}_j(k/\lambda)$.
Moreover, we find that the number of the bound states only depends on 
the  mass ratio $M/m$ and the coupling strength ratio $d/g$ (or $d_0/g_0$).  
This phenomenon is comprehensible due to
the scaling property of Hamiltonian (\ref{eqHamiltonian2D}).
Note that after the
scaling transition $x^\prime \to \lambda x$, $y^\prime \to \lambda y$,
the coupling constant $d$
and $g$ are rescaled as $g^\prime \to g/\lambda$, $d^\prime \to d/\lambda $
while the spectrum is rescaled
as $\epsilon_n^\prime \to \lambda^{-2} \epsilon_n$.
This scaling property keeps the structure of
the spectra invariant, thus the number of bound states remains constant
for fixed $M/m$ and $d/g$.
However, when the
three-body interaction is presented, this scaling property is broken. This will be
discussed in section IV.

\section{The phase diagram and emerged three-body bound states} \label{secPD}
\begin{figure}[tbp]
\centering
\includegraphics[width=0.5\textwidth]{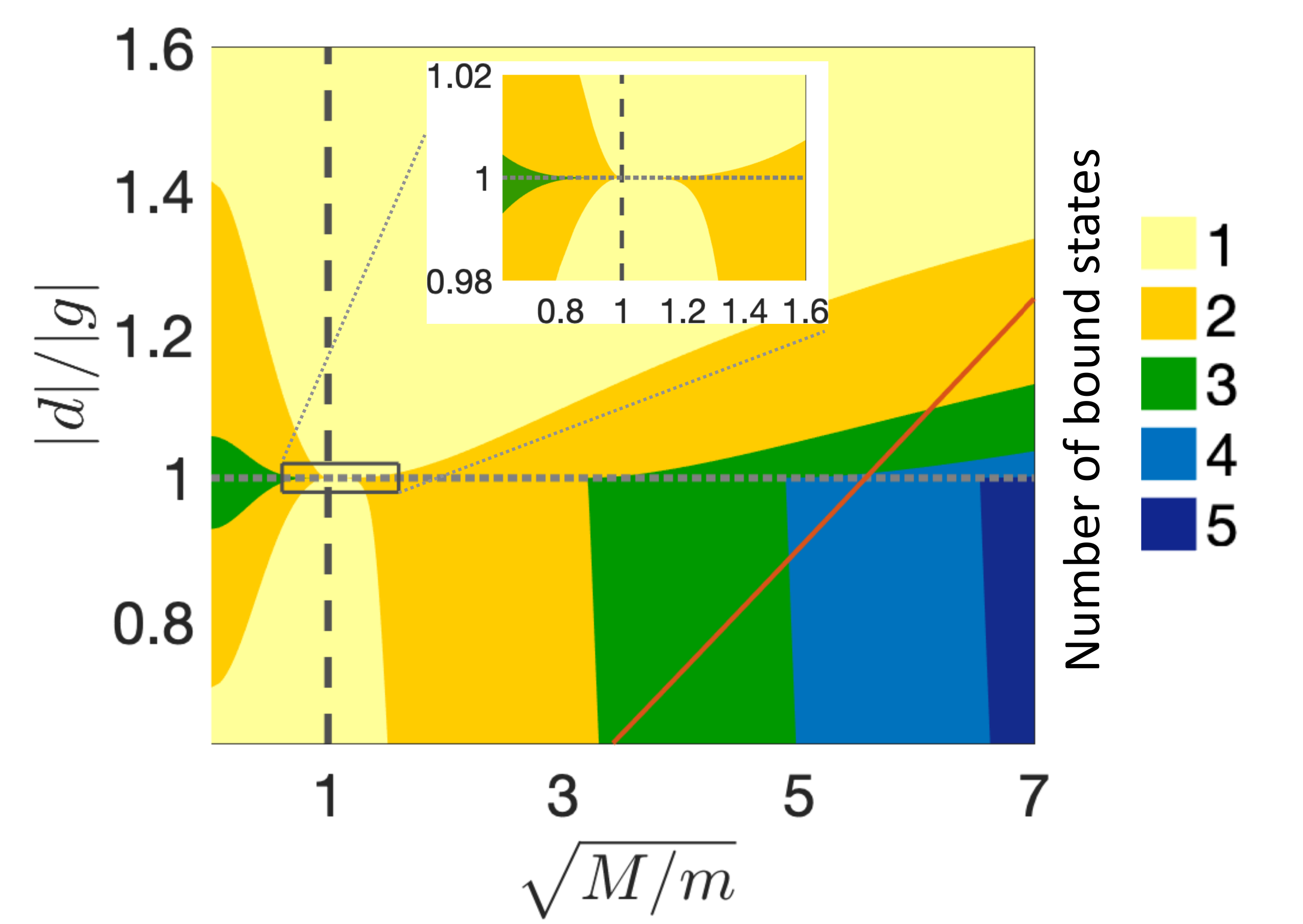}
\caption{The phase diagram of the number of three-body bound states.
The X-axis is the square root of the mass ratio, and the Y-axis
is the ratio of the absolute value of the coupling strength $d$ and $g$. Here we consider the attractions
with $d<0$ and $g<0$. Red line represents the relation between $|d|/|g|$
and $\sqrt{M/m}$ with $|g_0|/|d_0|=4$, where
$|d|/|g|=|d_0|/|g_0|\sqrt{1/2(1+M/m)}$.}
\label{fig2}
\end{figure}

Although the limit cases
$M/m \to \infty, {\vert}d{\vert}/{\vert}g{\vert}\to{0}\text{ or }\infty$
have been
discussed by Kartavtsev \textsl{et al.} \cite{Kartavtsev2009} and Mehta
\textsl{et al.} \cite{Mehta2014} under the one channel approximation and BOA. 
There are many regions of $M/m$ and $d/g$ 
still remain unexplored. By exactly solving the integral equation (\ref{eqSTM}),
we get the wave function and present
the full phase diagram of the number of bound states in the parameter space 
spanned by $\sqrt{M/m}$ and $|d|/|g|$, see Fig. \ref{fig2}.
In the region $M/m>1$, the number of bound states is in agreement with
the results in articles \cite{Mehta2014,Mehta2015}.
The phase diagram provides rich
information near the integrable point $|d|/|g|=1, M/m=1$.
It shows that in the equal mass case
there is always only one bound state. Near the integrable point the phase diagram is
sensitive to $M/m$: when $|d|/|g| = 1$, one more three-body bound
state will emerge even $M/m$ is slightly changed. When $M/m=1$, the
number of three-body bound states keeps constant with varying $|d|/|g|$.

\begin{figure}[htbp]
\centering
\includegraphics[width=0.5\textwidth]{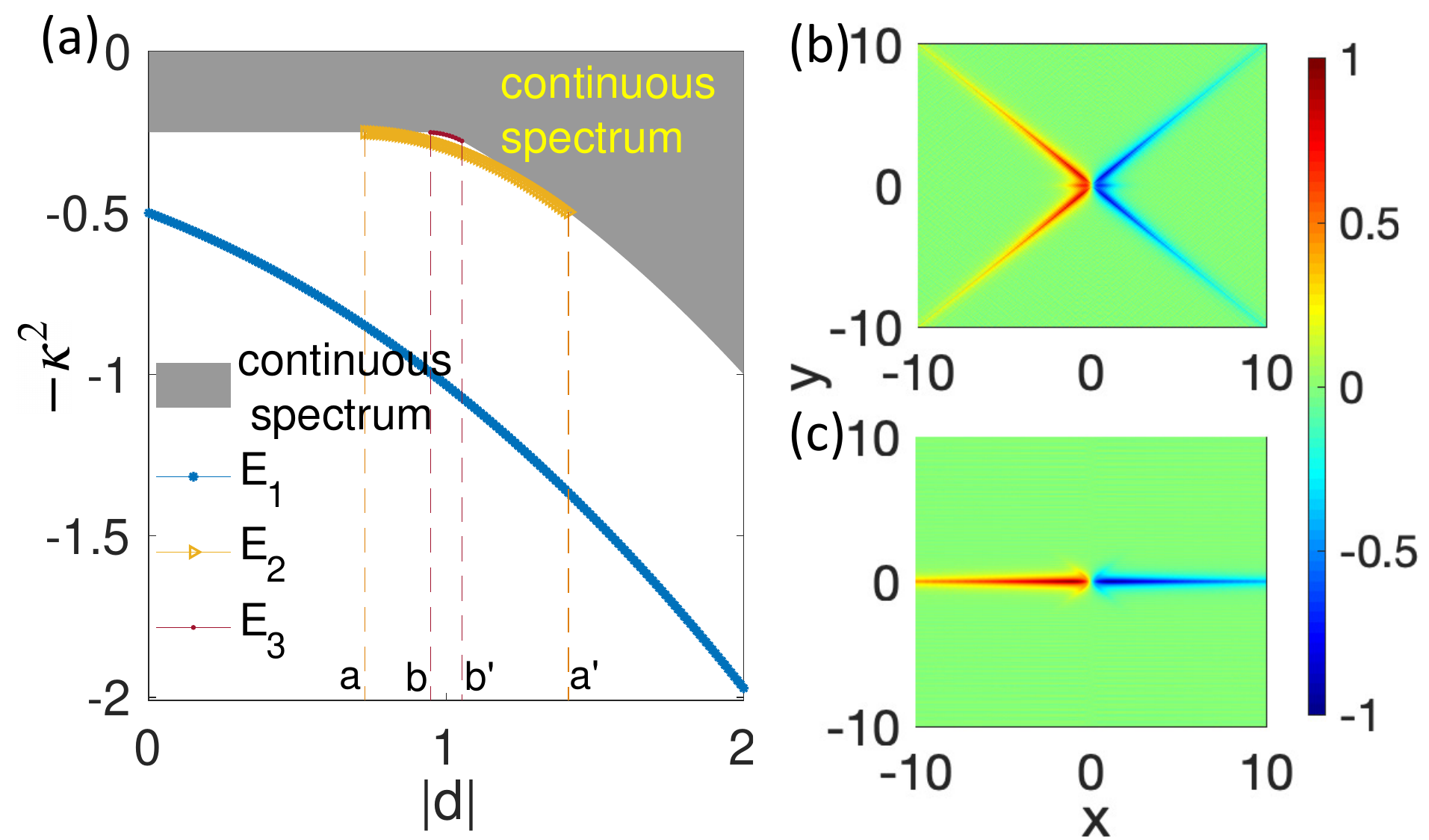}
\caption{(a) The binding energies of three-body bound states as function of
$d$ with $M/m=0.01$ and $g=-1$. The gray area denotes the continuous spectrum.
The orange dashed lines denote the critical values where the odd-parity
excited three-body bound state emerges at $|d|=a= 0.707$ and disappears at $|d|=a'=1.414 $, respectively.
The red dashed lines denote the critical values where the even-parity excited three-body
bound state emerges at $|d|= b=0.929$ and disappears at $|d|= b'=1.058$, respectively.
%
The wave functions of the odd-parity excitation states (the yellow line in (a)) near the
critical values (b) $\vert{d}\vert\to a$ and (c) $\vert{d}\vert\to a'$.}
\label{fig3}
\end{figure}

Moreover, in the HLL region $M/m<1$, Fig. \ref{fig2} shows the
emergence of extra excited three-body bound states near $g=d$.
To see it clearly, we present the binding energies of
three-body bound states as function of $d$ with $M/m=0.01$ and $g=-1$ in Fig. \ref{fig3}(a).
It is found that $3$ three-body bound states exist when $d/g \to 1$, one with
odd parity symmetry (the middle one) and two with even parity symmetry.

In the limit $M/m\rightarrow 0$ (or $\theta\rightarrow \pi/4$) with arbitrary parameters $g$ and $d$,
there exists Bethe ansatz solution for the odd-parity wavefunction of Hamiltonian (\ref{eqHamiltonian2D})
with energy $E= -\hbar^2(\kappa_1^2+\kappa_2^2)/(2\mu_{12,3})$:
\begin{equation} 
\psi(x,y) =
\left\{
\begin{aligned} \label{functionBA}
 &C\Big(\frac{d-\sqrt{2}g}{d-g/\sqrt{2}}e^{-\kappa_{1}x-\kappa _{2}y}
+\frac{g}{\sqrt{2}d-g} e^{-\kappa_{2}x-\kappa _{1}y} \\
&\quad \quad -e^{-\kappa _{2}x+\kappa _{1}y} \Big), \quad \quad \quad \text{for } 0<y<x;\\
 &C(e^{-\kappa _{1}x-\kappa _{2}y}-e^{\kappa _{1}x-\kappa _{2}y}),
\quad \text{for } y>\left\vert x \right\vert,
\end{aligned} 
\right.
\end{equation} 
where $\kappa _{1}=d/2-g/\sqrt{2}$, $\kappa _{2}=-d/2$ and $C$ is the normalization factor
of the wavefunction. The wavefunction in other regions can be obtained through the 
symmetries of wavefunction: $\psi (x,y) =\psi (x,-y) $ and $\psi (x,y) =-\psi (-x,-y) $.
The wavefunction in Eq. \eqref{functionBA} is confined in $y=0$ or $x=\pm y$,
which is not necessarily bounded in those directions. The existence of this bounded state puts
constraint conditions for $g$ and $d$. Along $y=0$, the wavefunction vanishes at 
$|x| \rightarrow \infty$, which gives us $\kappa _{1}>0$ and $\kappa _{2}>0$, thus, $d>\sqrt{2}g$.
Along $x \pm y=0$, the wavefunction vanishes at 
$|x \mp y| \rightarrow \infty$, which gives us $\kappa _{2}-\kappa _{1}>0$ 
and $\kappa _{1}+\kappa _{2}>0$, so $g/\sqrt{2}>d$. 
Put the pieces together, the odd-parity bound state exists within $g/\sqrt{2}>d>\sqrt{2}g$.
%
The even-parity bound states cannot be solved via Bethe Ansatz.
We show the energies of the bound states with $M/m=0.01$ in the Fig \ref{fig3}(a). 
The even-parity bounded state (marked by red line) emerges from the continuous spectrum from $|d|=0.929$ to
$|d|=1.058$. The numerical computation of the odd-parity bound state (marked by orange line) emerge from 
$|d|=0.707\approx 1/\sqrt{2}$ to $|d|=1.414\approx \sqrt{2}$, which is in agreement with the analytical result obtained
before.
Fig.\ref{fig3}(b) confirms that near $|d|=0.707$ one light particle is
combined tightly with the heavy one forming a molecule,
which is loosely combined with the other light particle.
Fig.\ref{fig3}(c) shows that the two
light particles form a dimer, which is loosely combined with
the heavy one at $|d|=1.414$.
Fig.\ref{fig3}(b) and (c) also have something to say about the threshold of atom-dimer continuous spectra.
When $|d|<1$, $|d|<|g|$, the trimer state exhibits 
near threshold that bound one heavy particle and one light particle tightly, as shown in Fig.\ref{fig3}(b), which suggests the dimer 
at the bottom of the threshold is formed by a heavy particle and a light particle.
When $|d|>1$, $|d|>|g|$, the dimer at the bottom of the threshold consists of two light particles, which can be inferred from 
Fig.\ref{fig3}(c) with similar reasoning.

\section{The effects of three-body interaction} \label{secTE}

It has been discussed in the end of section \ref{secMM} that the scaling property of the
Schr\"odinger equation (\ref{eqHamiltonian2D}) results in
the structure of the spectra relying only on $M/m$ and $d/g$.  However,
the three-body attraction breakes this scaling property and may cause significant
consequences \cite{Nishida2018,Sekino2018,Guijarro2018,Cheiney2018}. The three-body zero-range
interaction can be introduced by adding the term
$\hat{H}^{(3)} = t_0 \delta(x_3-x_1/2-x_2/2)\delta(x_1-x_2)$ in (\ref{eqHamiltonian}),
which corresponds to adding the term
\begin{align}
\hat{H}_{\text{red}}^{(3)} = t_B\delta(x)\delta(y)
\label{eqH3}
\end{align}
to the Schr\"odinger equation (\ref{eqHamiltonian2D}), where $t_B= \frac{2 t_0}{\sqrt{2M/m+1}}$.
Similar to (\ref{eqSchEq}), the Schr\"odinger equation in momentum space is given by
\begin{align}
-(p_x^2+p_y^2+\kappa^2) \frac{u(p_x,p_y)}{c_3} - \sum_{i=1}^3 \frac{g_i}{2\pi} \frac{f_i(d_i)}{c_3}= 1,
\label{eqSchEq3}
\end{align}
where $g_i$, $d_i$ and $f_i$ are the same as defined in section \ref{secMM},
and
\begin{align}
c_3 = \frac{t_B}{4\pi^2}\int_{-\infty}^{\infty} \mathrm{d}k_x\mathrm{d}k_y u(k_x,k_y)
\label{threec3}
\end{align}
represents the
three-body interaction in momentum space.
However, the wave function $u(k_x,k_y)$ in momentum space is proportional to
$-c_3/(p_x^2 + p_y^2 + \kappa^2)$. Equation \eqref{threec3} experiences logarithmic 
divergence when integrating in the whole momentum space, which suggests the requirement 
of renormalization procedure for the three-body interaction strength $t_B$.

Based on Schr\"odinger equation (\ref{eqSchEq3}) and the same method developed in
section \ref{secMM}, the STM equations with three-body interaction is obtained by
adding  an extra term ${-c_3\pi}/{\sqrt{k^2 + \kappa^2}}$ into (\ref{eqSTM}), which gives rise to
\begin{align}
&\left(1+\frac{g_i}{2\sqrt{k^2+\kappa^2}} \right)F_i(k) = -\frac{\pi}{\sqrt{k^2 + \kappa^2}}\notag \\
&+\sum_{j \neq i} \int \frac{\mathrm{d} k'}{2\pi} \frac{- g_j \vert \sin(\theta_i - \theta_j)\vert F_j(k')}
{k'^2+k^2-2 k k'\cos(\theta_i-\theta_j) + \kappa^2\sin^2(\theta_i-\theta_j)},
\label{eqSTMthreebody}
\end{align}
where $F_i(k)=f_i(k)/c_3$. For fixed $g_i$ and $M/m$, with any real $c_3$, 
equation (\ref{eqSTMthreebody}) gives unique $F_i(k)$.
In solving $F_i(k)$, we encounter the matrix $(\bm{M}(E)-\bm{I})$,
which is invertible in the bound state sector except for the isolate points \cite{Weinberg1964},
which corresponds to the
discrete bound energies in (\ref{eqSTM}).

The solutions of equations
(\ref{eqSTM}) and (\ref{eqSTMthreebody}) are different in
ultraviolet behaviours, it can be proved that $f_i(k)\propto 1/k^2$ in (\ref{eqSTM})
and $F_i(k) \approx -\pi/\sqrt{k^2+\kappa^2}$ in (\ref{eqSTMthreebody}), whose derivations are given in
appendix A.
Since a function which decays faster at large momentum is preferable in numerical
methods,
it is convenient to introduce the substitution of $F_i(k)$ by
$F_i(k)=-\pi/\sqrt{k^2+\kappa^2}+h_i(k)$, here $h_i(k) \propto 1/k^2$ at $k \to \infty$.
Taking this relation into (\ref{eqSTMthreebody}), the integral equation
for $h_i(k)$ is obtained
\begin{align}
&\left(1+\frac{g_i}{2\sqrt{k^2+\kappa^2}} \right)h_i(k) =
-\pi\sum_{j}g_j \eta(k,\kappa^2,\vert{\theta_i-\theta_j}\vert)
 \notag \\
&+\sum_{j \neq i} \int \frac{\mathrm{d} k'}{2\pi} \frac{- g_j \vert \sin(\theta_i - \theta_j)\vert h_j(k')}
{k'^2+k^2-2 k k'\cos(\theta_i-\theta_j) +\kappa^2\sin^2(\theta_i-\theta_j)}
\label{eqSTMthreebody2}
\end{align}
with analytic function $\eta(k,\kappa^2,\theta)$ defined by
\begin{align}
&\eta(k,\kappa^2,\theta) =
\frac{-\frac{1}{2\pi}\vert{\sin\theta}\vert}{\sqrt{k^2+\kappa^2}(k^2+\kappa^2\cos^2\theta)}
\notag \\
&\times\left(2\vert{k}\vert\text{arcoth}\frac{\sqrt{k^2+\kappa^2}}{\vert{k}\vert}
+\frac{(\pi-2\theta)\sqrt{k^2+\kappa^2}}{\tan\theta}\right)
\notag
\end{align}
and $\eta(k,E,0) \triangleq\lim_{\theta\rightarrow 0}\eta(k,E,\theta)=-(2k^{2}+2\kappa^{2})^{-1}$.

Substituting equations (\ref{eqSchEq3}) into (\ref{threec3}), we can get
\begin{align}
\frac{1}{t_B}&=
-\frac{1}{4\pi}\ln\frac{\Lambda^2+\kappa^2}{\kappa^2} \notag \\
&-\frac{1}{4\pi^2}\sum_i\int_{\Lambda} \frac{\mathrm{d}S}{2\pi}
\frac{g_i  F_i(d_i(p_x,p_y))}
{p_x^2+p_y^2+\kappa^2},
\label{eqRG2}
\end{align}
where $\int_\Lambda \mathrm{d} S \triangleq
\int_{p_x^2+p_y^2<\Lambda^2} \mathrm{d}p_x\mathrm{d}p_y$ is the
two dimensional integral with cut-off $\Lambda>0$. We apply the
momentum-cutoff regularization scheme \cite{Thorn1979,Camblong2002}
with momentum cut-off $\Lambda$.

The relation between the bare coupling constant $t_B$ and the renormalized
coupling constant $t_R$ can be written as
\begin{align}
\frac{1}{t_B} = \frac{1}{t_R} - \frac{1}{4\pi}\ln{\frac{\Lambda^2}{\mu^2}},
\label{eqRG}
\end{align}
where $\mu^2$ is the emerged energy scale. The renormalized coupling constant
$t_R$ is obtained by cutting off the logarithmically divergent part 
in $t_B$ with scaling $\mu$ \cite{Coleman1973,Faddeev2006}.

Substituting equation (\ref{eqRG}) into (\ref{eqRG2}) and replacing $F_i(k)$ with
$-\pi/\sqrt{k^2+\kappa^2}+h_i(k)$ in equations (\ref{eqRG2}), we arrive at the
relation between the solution $h_i(k)$, the renormalized coupling constant $t_R$
and energy scale $\mu^2$:
\begin{align}
\frac{1}{t_R} = \frac{1}{4\pi}\ln\frac{\kappa^2}{\mu^2}
+ \sum_i \frac{g_i}{8}\left(\frac{1}{\kappa}
- \frac{1}{\pi^2}\int \mathrm{d} k \frac{h_i(k)}{\sqrt{k^2 + \kappa^2}}\right) .
\label{eqTR}
\end{align}
The equations (\ref{eqSTMthreebody2}) and (\ref{eqTR}) completely determine the
relation between $g_i,\theta_i,\mu,t_R$ and $\kappa^2$.
Begin with one parameter $t_B$, the renormalization scheme introduces
two quantities
$t_R$ and $\mu$  for the three-body interaction. The physical three-body coupling strength the particles feel is
$t_R$.  However, since we can choose
$t_R$ arbitrarily, the renormalized three-body interaction can be described by
one scaled parameter only \cite{Camblong2001,Camblong2002}. To this end we introduce the three-body
scattering length $a_3$ which describes the
asymptotic behaviour of the wave function $\Psi=\psi/c_3 \propto
\ln\frac{\rho}{a_3}$ when $\rho=\sqrt{x^2+y^2} \to 0$ \cite{Guijarro2018}.
To obtain $a_3$ we need to expand the wave function
at $\rho \to 0$:
\begin{align}
\Psi(\rho) = - K_0(\vert{\kappa\rho}\vert)
+ 2\pi (\frac{1}{t_R}-\frac{1}{4\pi}\ln\frac{\kappa^2}{\mu^2}),
\label{eqAsymp}
\end{align}
which can be obtained by solving $u(p_x,p_y)/c_3$ from (\ref{eqSchEq3}) and transforming
it to coordinate space,
where  $K_0(x)$ is the modified Bessel function of the second kind and has the
asymptotic behavior
 $K_0(\vert{x}\vert) \approx -\ln\vert\frac{x}{2}\vert-\gamma$
at $\vert{x}\vert\to{0}$ with Euler
constant $\gamma \approx 0.57722$. Substituting it into (\ref{eqAsymp}),
we arrive at the close form of $a_3$:
\begin{align}
-\ln\frac{a_3}{2} = \gamma + \frac{2\pi}{t_R} + \ln\mu,
\label{eqa3}
\end{align}
where $a_3>0$ for the RHS of \eqref{eqAsymp} being real. The equations (\ref{eqSTMthreebody2}), (\ref{eqTR}) and (\ref{eqa3}) give the relation
between two-body scattering lengthes and three-body scattering length.
In solving the equations (\ref{eqSTMthreebody2}), (\ref{eqTR}) and \eqref{eqAsymp},
the same difficulty arises as in Eq.(\ref{eqSTM}): it is not easy to solve $\kappa^2$ for a given $a_3$.
The opposite is undemanding, and we handle this difficulty in the
same manner as in section \ref{secMM}, $\mathrm{i.e.}$, we search for the three-body
scattering length $a_3$
for given $\kappa^2$ and two-body coupling constants $d$ and $g$.

We have discussed in section \ref{secPD} that the system
with only two-body interactions can exhibit multi three-body bound states.
It has been proved that the three-body interaction alone can exhibit one three-body
bound state with energy $E=-\hbar^2\kappa^2/(2\mu_{12,3})=-4\hbar^2e^{-2\gamma}/(2a_3^2\mu_{12,3})$ \cite{Camblong2002,Sekino2018}.
Refs. \cite{Guijarro2018} and \cite{Nishida2018} showed that when $M=m$, $d_0=g_0<0$,
one more three-body bound state would emerge once the
three-body interaction is introduced.
Now an interesting question arises: for the mass imbalenced system with attractive 
two-body interaction, does the system exhibits additional three-body bound states with arbitrarily tuned three-body interaction?
\begin{figure}[t]
\centering
\includegraphics[width=0.45\textwidth]{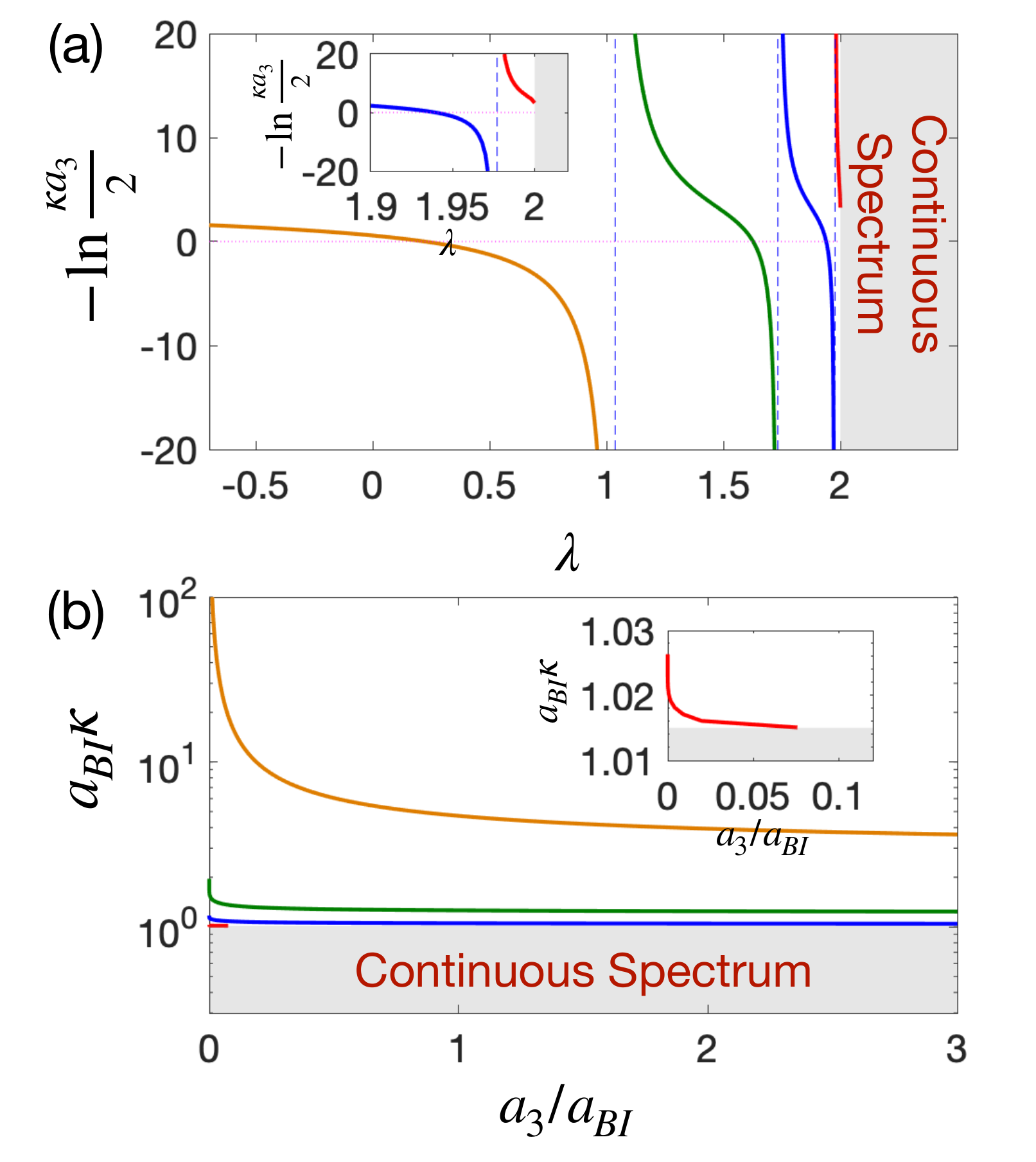}
\caption{The binding energies for the system with $M/m=16$
and $g_0/d_0=4$ in the presence of three-body interaction. (a) The relation between $-\ln\frac{\kappa{a_3}}{2}$ and $\lambda$,
 where $a_3$ is three-body scattering length and $\lambda=-g/\kappa$ ($\kappa$ is regarded as unit for two-body and three-body
scattering length.). So the two-body coupling
constant $d$ is then expressed as
$d=-\lambda\kappa\frac{d_0}{g_0}\sqrt{\frac{m+M}{2m}}$ (refer to
relation (\ref{eqGdtheta})). The threshold $E_{\text{th}} = -\hbar^2(g/2)^2/(2\mu_{12,3})$  
also corresponds to $\lambda=2$ ($\kappa=-g/2$).
(b) The binding energy related parameter $a_{BI}\kappa$ (the binding energies $E= -\hbar^2\kappa^2/(2\mu_{12,3})$)
as functions of the three-body to two-body BI scattering length ratio $a_3/a_{BI}$.
The grey area represents the atom-dimer scattering continuum.}
\label{fig4}
\end{figure}
The answer is negative. In certain parameter regions, there is no additional three-body bound state, 
as one can see from Fig. \ref{fig4}.
In Fig. \ref{fig4} we demonstrate three-body bound states with $M/m=16$
and $g_0/d_0=4$.  For given $\lambda=-g/\kappa$, the three-body scattering length 
is determined uniquely. As $\lambda$ increases, $-\ln\frac{\kappa{a_3}}{2}$ repeatedly runs from 
$+\infty$ to $-\infty$ monotonically and continuously,  which is shown in Fig. \ref{fig4}(a).
Figure. \ref{fig4} (b) can be obtained from a coordinate transformation of Fig. \ref{fig4}(a).
When $a_3/a_{BI}\rightarrow \infty$ ($t_R=0$),
there remain $3$ three-body bound states, which are induced only by two-body interaction.  
For arbitrary three-body interaction, those $3$ states always exist.
For $a_3/ a_{BI}<0.076$, an additional three-body bound state emerges out of the atom-dimer continuum,
which is induced by three-body interaction. This puts an upper bound for $a_3/ a_{BI}$ as $a_{max,BI}=0.076$, 
below which there exists the additional three-body bound state.

\begin{figure}[t]
\centering
\includegraphics[width=0.45\textwidth]{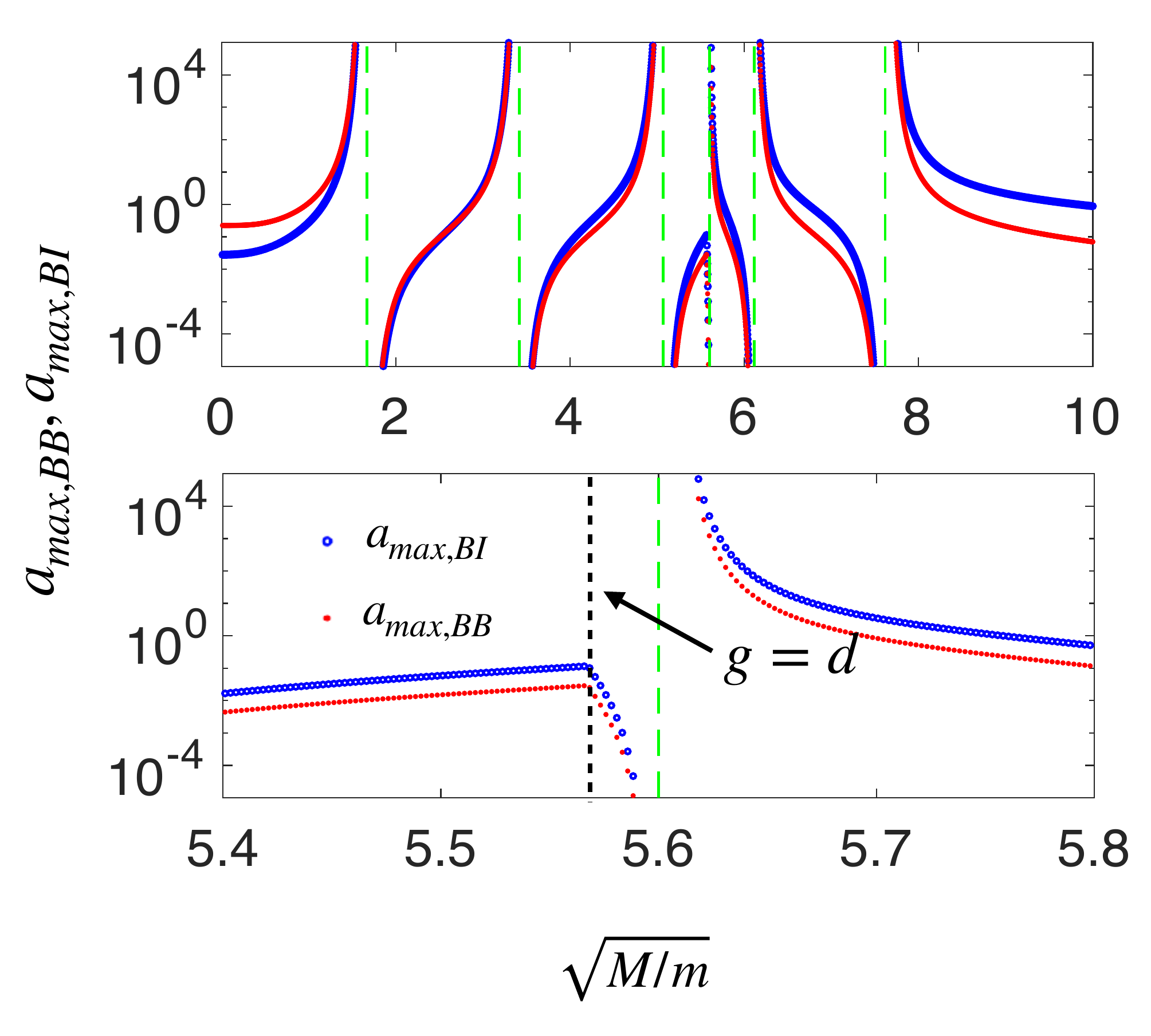}
\caption{The maximum three-body scattering length
in units of two-body BB (BI) scattering length $a_{BB}$($a_{BI}$)
v.s. the square root of mass ratio $\sqrt{M/m}$ at
the two-body threshold for the system with the two-body coupling $g_0/d_0=4$.
The dotted line is $\sqrt{M/m}=5.57$, which corresponding to $g=d$.
 The  area near $\sqrt{M/m}=5.57$
is zoomed in in the bottom panel.}
\label{fig5}
\end{figure}

From the analysis above, we conclude that the three-body interaction can
not always bring one more three-body bound state, which depends on the
mass ratio, the three-body and two-body scattering lengths.
In Fig. \ref{fig5}, we plot the maximum three-body scattering length
in units of two-body BB (BI) scattering length $a_{BB}$ ($a_{BI}$) as a function of
the square root of mass ratio $\sqrt{M/m}$ with fixed $g_0/d_0=4$.
Altering mass ratio $\sqrt{M/m}$ can change  $g/d$. The relation between $\sqrt{M/m}$
and $g/d$ is shown in Fig. \ref{fig2}  as red line.
The numerical result shows that there are some singularities, which occur at green dashed lines, as
shown in the Fig. \ref{fig5}. 

As we can see from Fig. \ref{fig5}, by increasing $\sqrt{M/m}$ from 
$\sqrt{M/m}=4$ (the case in Fig. \ref{fig4}),  $a_{\text{max},BI,BB}$ increases until 
it meets infinity at $\sqrt{M/m}=5.06$, as a consequence of which, the intersection value of the three-body bound 
state and the atom-dimer continuum goes from a certain value to infinity. In this sense, 
at $\sqrt{M/m}=5.06$, there are $4$ three-body bound states for arbitrary $a_3$, among which the one
with the smallest $\kappa$ is induced by three-body interaction. This statement can be verified by comparing
with Fig. \ref{fig2}. The system without three-body interaction meets transition point from $3$ to $4$ three-body bound states 
at $\sqrt{M/m}=5.06$ and $|d_0|/|g_0|=4$.
Continuously increase $\sqrt{M/m}$ until $5.57$, there are $5$ three-body bound 
states at presence with $a_{\text{max},BB,BI}$  increases from $0$ (at the exact point 
$a_{\text{max},BB,BI}=0$ there are actually four three-body bounded states induced by two-body interaction, 
which can also be verified by comparing with Fig. \ref{fig2}) to a certain value. Among the $5$ three-body 
bound states the one with the smallest $\kappa$ is induced by three-body interaction. 
At $\sqrt{M/m}=5.57$, we have $d/g=1$. Increase $\sqrt{M/m}$ again, $a_{\text{max},BB,BI}$ 
begins to decrease. This turning point is non-smooth, which is a result of the unsmoothly change of threshold 
from $\sqrt{M/m}<5.57$ ($d/g<1$) to $\sqrt{M/m}>5.57$ ($d/g>1$). With $\sqrt{M/m}<5.57$ and $\sqrt{M/m}<5.57$, the thresholds are $E_{\text{th}} = -\hbar^2(d/2)^2/(2\mu_{12,3})$ and $E_{\text{th}} = -\hbar^2(g/2)^2/(2\mu_{12,3})$,
of which the first-order derivatives are discontinued at $\sqrt{M/m}=5.57$.
The reason for $a_{\text{max},BB,BI}$ is a monotonically increasing (decreasing)
function of the mass ratio for $\sqrt{M/m}<5.57$ ($\sqrt{M/m}>5.57$) is that the number of the three-body bound states
induced only by two-body interaction
is increasing (decreasing) at $\sqrt{M/m}<5.57$ ($\sqrt{M/m}>5.57$), see Fig. \ref{fig2} for reference.
Keep increasing $\sqrt{M/m}$ to $5.61$, $a_{\text{max},BB,BI}$ decreases 
to $0$, which suggests the vanishment of one three-body bound state. At $\sqrt{M/m}=5.61$, 
there remains only four three-body bound states, which are all resulted from two-body interaction. 
Again increase $\sqrt{M/m}$ until $6.1$, $a_{\text{max},BB,BI}$ decreases from $\infty$ to 
zero. In this interval, we have $3$ three-body bound states induced by two-body interaction.

An interesting fact is that the locations of the green dashed lines in Fig. \ref{fig5} 
match exactly with the intersection of the red line and the phase boundaries in Fig. \ref{fig2}. 
This can be understood as when $a_{\text{max},BB,BI}=0$, the energy of three-body bound state
induced by two-body interaction with lowest $\kappa$ approaches to the 
atom-dimer continuum spectrum at $a_3\rightarrow\infty$, where the particles experience no 
three-body interaction. This is exactly the condition for the transition of the number of two-body interaction
inducing three-body bound states to occur without three-body interaction. 
The explanation above also works for why there is 
always an additional three-body bound state in mass balanced case.
The intersection of the dashed line ($M/m=1$) and dotted line ($|d|/|g|=1$) in Fig. \ref{fig2} is a transition point when varying 
the mass ratio along an interval containing $M/m=1$ with fixed $|d|/|g|$.

\section{Summary}

In summary, we studied the bound states of a 1D three-body mass-imbalanced system with two-body attractive interaction.
In the absence of three-body interaction, we presented the phase diagram of the number of three-body bound states
by solving the STM equations with arbitrary $d/g$ and $M/m$. We developed some computational
techniques and applied them to obtain the complete phase diagram. We demonstrated that the LLH system 
has at most three three-body bound states. Particularly, in the limit of $M/m \rightarrow 0$ the LLH system has the
Bethe Ansatz solution, which further verifies the validity of our results. Moreover,
we found that the presence of the three-body interaction may lead to one more bound state. However,
this additional three-body bound state would not always exist, but depends on the mass ratio and the 
ratio of coupling strength $d_0/g_0$. The existence of the additional three-body bound state is independent of
the three-body interaction at some special parameter points which correspond to the transition points of the 
number of three-body bound states induced solely by two-body attractive interaction.

The techniques to solve the STM equations may be applied to study mass-imbalanced
four-body or $N$-body system. Our results may help understanding of how 
mass-imbalanced particles are bounded with two-body attractive interactions and three-body interaction.

\begin{acknowledgments}
The work is supported by NSFC under Grants No.11974413, the National Key
Research and Development Program of China (2016YFA0300600) and the Strategic Priority Research Program of Chinese Academy of Sciences under Grant No. XDB33000000. Y.-C. Yu was supported by the National Science Foundation (NSF) for
Young Scientists of China under Grant No.11804377.
\end{acknowledgments}

 \appendix
\section{The large momentum behavior of solutions of Eq. \eqref{eqSTM} and Eq. \eqref{eqSTMthreebody} }

Equation \eqref{eqSTM} can be written as
\begin{equation}
f_i(k)=\sum_{j \neq i}\int dk' f_j (k') G_{i,j}(k,k'), \label{eqfg}
\end{equation}
where 

\begin{align}
&G_{i,j}(k,k') = \frac{1}{\left(1+\frac{g_i}{2\sqrt{k^2+\kappa^2}} \right)}  \notag \\
& \times \sum_{j \neq i} \int \frac{\mathrm{d} k'}{2\pi} \frac{- g_j \vert \sin(\theta_i - \theta_j)\vert }
{k'^2+k^2-2 k k'\cos(\theta_i-\theta_j) +\kappa^2\sin^2(\theta_i-\theta_j)}. \label{eqST}
\end{align}

Expand $G_{i,j}(k,k')$ at $1/k \rightarrow 0$,

\begin{align}
&G_{i,j}(k,k') =  \frac{- g_j \vert \sin(\theta_i - \theta_j)\vert}{k^2}+ o(\frac{1}{k^3}).  \label{eqg}
\end{align}
$f_i(k)$ at large momentum is
\begin{align}
f_i(k)= \frac{A_i}{k^2}+  o(\frac{1}{k^3}), \label{EQFK}
\end{align}
where $A_i=-\sum_{j \neq i} g_j \vert \sin(\theta_i - \theta_j)\vert $. So, $f_i(k) \propto \frac{1}{k^2}$ in ultraviolet region.

Similarly, Eq. \eqref{eqSTMthreebody} can be written as
\begin{align}
F_i(k)&=- \frac{1}{\left(1+\frac{g_i}{2\sqrt{k^2+\kappa^2}} \right)} \frac{\pi}{\sqrt{k^2 + \kappa^2}} \notag \\
&+ \sum_{j \neq i}\int dk' F_j (k') G_{i,j}(k,k'), \label{eqfg}
\end{align}
By large momentum expansion,
\begin{align}
F_i(k)+\frac{\pi}{\sqrt{k^2 + \kappa^2}} = \frac{A_i+\frac{\pi g_i}{2}}{k^2}+  o(\frac{1}{k^3}). \label{EQFK}
\end{align}
Thus, $F_i(k) \approx -\frac{\pi}{\sqrt{k^2 + \kappa^2}}$ and $F_i(k)+\frac{\pi}{\sqrt{k^2 + \kappa^2}} \propto \frac{1}{k^2} $  in ultraviolet region.


\end{document}